\documentclass[twocolumn,showpacs,pra,aps,10pt]{revtex4-1}
\usepackage{amsmath}
\usepackage{epsfig}
\usepackage{graphicx}
\usepackage{epstopdf}
\usepackage{footnote}
\usepackage{xcolor}

\begin{document}

\title{Simple criteria for noise resistance of two qudit entanglement}
\author{Arijit Dutta, Junghee Ryu, Wies{\l}aw Laskowski, and Marek \.Zukowski}
\affiliation{Institute of Theoretical Physics and Astrophysics, University of Gda\'{n}sk, 80-952 Gda\'{n}sk, Poland}

\begin{abstract}Too much noise kills entanglement. This is the main problem in its production and  transmission.
We use a handy approach to indicate noise resistance of  entanglement of a bi-partite system described by $d\times d$ Hilbert space.
Our analysis uses a geometric approach based on the fact that if a scalar product of a vector $\vec{s}$ with a vector $\vec
{e}$ is less than the square of the norm of $\vec{e}$, then $\vec{s}\neq\vec{e}$. We use such concepts for correlation tensors of separable and entangled states. As  a general form   correlation tensors for pairs of qudits, for $d>2$, is very difficult to obtain, because one does not have a Bloch sphere for pure one qudit states, we use a simplified approach. The criterion reads: if the largest Schmidt eigenvalue of a correlation tensor is smaller than the square of its norm, then the state is entangled. this criterion is applied in the case of various types of noise admixtures to the initial (pure) state. These include 
 white noise, colored noise,  local depolarizing noise and amplitude damping noise. A broad set of numerical and analytical results is presented. As the other simple criterion for entanglement is violation of Bell's inequalities,  we also find critical noise parameters  to violate specific family of Bell inequalities (CGLMP), for maximally entangled states. We give analytical forms of our results  for $d$ approaching infinity.
\end{abstract}

\maketitle

\newcommand{\bra}[1]{\langle #1\vert} 
\newcommand{\ket}[1]{\vert #1\rangle} 
\newcommand{\abs}[1]{\vert#1\vert} 
\newcommand{\avg}[1]{\langle#1\rangle}
\newcommand{\braket}[2]{\langle{#1}|{#2}\rangle}
\newcommand{\commute}[2]{\left[{#1},{#2}\right]}

%------------------------------------------------------------
\section{Introduction}
%------------------------------------------------------------
Detection of entanglement is a fundamental problem in quantum information .  Many theoretical works have been done to establish a criterion to identify multipartite entanglement \cite{Horodecki09, Guhne09, Vicente, Laskowski11, Laskowski15, Huber2010, Spengler}. Although not all of these works have  direct impact on experiments, still they provide a  mathematical background to understand the phenomena. 

An arbitrary density operator for two qudits can be written as
\begin{eqnarray}
\label{eq:2d_state}
\rho&=&\frac{d-1}{2d}\Big( M_{0}\otimes M_{0}
+\sum_{i, j=1}^{d^2-1}T_{ij}M_{i}\otimes M_{j}\\
&+&(d-1)\sum_{i=1}^{d^2-1}(T_{0i}M_{0}\otimes M_{i}
+T_{i0}M_{i}\otimes M_{0})\Big), \nonumber
\end{eqnarray}
where  $ T_{ij}, T_{0i}, T_{i0}$ are components of the  correlation tensor $\hat{T}, $ and $M_{i}(\text{except}~i=0)$  are the generalized Gell-Mann matrices (see Appendix~\ref{apx:GM_matrix} for details) with $M_0 = \sqrt{\frac{2}{d(d-1)}}\openone_d.$ The components $ T_{ij}$ are real and  given by $T_{ij}= c(d) \textrm{Tr} [\rho (M_{i} \otimes M_{j})]$, where the coefficient $c(d)$ depends  on the dimensionality (see Appendix~\ref{B}).

In ~\cite{Badziag08} it  has been shown that the correlation tensor could be utilized to construct  entanglement indicators.  The  nonlinear character of the resulting entanglement indicator of~\cite{Badziag08} make them more versatile than linear entanglement witnesses. The study of ~\cite{Badziag08} was mainly done for multi-qubit systems. Here,  we use  the geometric approach of ~\cite{Badziag08} to detect entanglement for pairs of systems,  more complicated than qubits. In our approach  we do not study the dynamics of noise, but we concentrate on  properties of the  state after evolving through a noisy channel. The white noise usually arises due to imperfections in experimental set up, while a ``colored''  noise can appear when multi particle entanglement is produced via multiple emissions ~\cite{SEN}. We also analyze the effects of different local noises,  modeling random environment and dissipative process~\cite{Nielsen_book}.

We  find that  in some of the cases there exist significant differences in numerical values of the critical noise admixture  parameters for having  entanglement after a transfer via various  noisy channels. A study of this kind was done for qubits  in Ref.~\cite{Laskowski10}. Here,  we study  higher dimensions. Note that,  a  different approach to obtain the criterion of separability of entangled qutrits in noisy channel  is given   in Ref.~\cite{Chcińska}. 

Another way to detect entanglement is to violate Bell inequalities~\cite{Bell64}. In our paper,  we discuss resistance to noise of   Bell-type inequalities for arbitrary dimension $d,$  e.g.  Collins-Gisin-Linden-Massar-Popescu (CGLMP) inequality~\cite{CGLMP}.

%------------------------------------------------
\section{Simplified entanglement condition}
\label{ct}
%------------------------------------------------
 Separable states can be described by a fully separable extended correlation tensor, $\hat{T}^{\text{sep}}=\sum_i p_i\hat{T}_i^{\text{prod}}$, where $\hat{T}_i^{\text{prod}}=\hat{T}_i^1\otimes\hat{T}_i^2$, and each $\hat{T}_{i}^{k}$ is  the  $d^2$-dimensional  generalized Bloch vector of $k$th single qudit subsystem. To produce  entanglement conditions  we employ a geometrical approach based on correlation functions, which was introduced in Ref.~\cite{Badziag08}. The criterion has the following form: the state  $\rho$  described by its correlation tensor $\hat{T}$ is entangled, if 
\begin{equation}
\max_{\hat{T}^{\text{prod}}} (\hat{T}, \hat{T}^{\text{prod}}) < (\hat{T}, \hat{T}) = ||\hat{T}||^{2}.
\label{eq:ent_criterion0}
\end{equation}
 The scalar product may be  defined in various ways, provided it satisfies all required axioms. However the simplest choice is :
\begin{equation}
\label{semid}
(\hat{X}, \hat{Y})=\sum_{i, j=1} ^{d^2-1}X_{ij} Y_{ij}.
\end{equation}
In such a case, the maximum of the left-hand side of Eq.~($\ref{eq:ent_criterion0}$) can be obtained by the highest generalized Schmidt coefficient of the correlation tensor $\hat{T}$ as
\begin{equation}
T_{\text{max}}=\max_{\vec{m}_1, \vec{m}_2} (\hat{T}, \vec{m}_1\otimes \vec{m}_2) ,
\label{eq:ent_criterion1}
\end{equation}
where $\vec{m}_k$ is a $(d^2-1)$-dimensional generalized Bloch vector which describes a quantum state of $k$th subsystem. Note  that, for $d>2$ we do not have a Bloch sphere. Thus $\vec{m}_k$'s represent the admissible Bloch vectors which represent a physical state. One can also take the maximum of the left-hand side of~($\ref{eq:ent_criterion0}$) as the square root of the  highest eigenvalue of $TT^\dagger$, denoted by $L_{\text{max}}$. However, note that 
$L_{\text{max}}\geq T_{\text{max}},$ and this estimate gives a less robust criterion for entanglement. Still it is technically much easier, and thus we shall use it.

%---------------------------
\section{Basic aims}
\label{state}
%---------------------------
We analyze entanglement of bipartite quantum states which are initially pure. For technical reasons  we always put  them in the Schmidt form:
\begin{equation}
\ket{\psi} = \sum_{i=0}^{d-1} c_i \ket{ii}.
\label{state0}
\end{equation}
We study  various noisy channels, i.e., white, product, colored, local depolarizing and amplitude damping. In some cases, one can define a critical parameter $v_{\text{crit}}$ for the noisy output, of the form $\rho(v)=v \ket{\psi} \bra{\psi} + (1-v)\rho_{\text{noise}}$, such that for $v > v^d_{\text{crit}, ENT}$ the state $\rho(v)$ is entangled. The action of a channel on a density matrix of a two-qudit can also  be written in terms of its operation elements, more precisely $\rho\to\sum_{u, v} E_u(p)\otimes E_v(p)\rho (E_u(p)\otimes  E_v(p))^{\dagger} $, where $E_l$  is an operation element of the channel parametrized by $p,$  a certain strength parameter to be specified later. We define  this parametrization  in such a way, so that $p=1$ defines no noise, just unitary(local) evolution, while $p=0$ signifies the maximally noisy state  leaving the channel. Hence,  one can show presence of entanglement by satisfying the condition $p > p^d_{\text{crit}, ENT}.$ Thus such a description is a generalization of the one involving $v_{\text{crit}}.$ We introduce a  quantity 
\begin{equation}
\label{square}
\xi(p^d_{\text{crit}, ENT})=\text{min}\left(\left(\frac{||\hat{T}(p^d_{\text{crit}, ENT})||^2}{||\hat{T}(p=1)||^2}\right)^{\frac{1}{2}}, 1\right)
\end{equation}
to compare different noisy channels on the same footing. In order to check  the entanglement condition (\ref{eq:ent_criterion0}), we  put  the correlation tensor of a  state $\eqref{state0}$   in a Schmidt form itself. The general form of the tensor elements  is given in Appendix~\ref{A}. Note that the Schmidt form of the state (\ref{state0}) does not imply a diagonal form of the corresponding tensor. Even for a two-qutrit state non-diagonal elements appear. This is not the case for qubits. Because of,  the above reason, calculations of $L_{\text{max}}$ require the diagonalization of the tensor, which is a computationally difficult problem even for small $d$.

For white, product, and colored noises  we analytically compute the critical parameter $v^d_{\text{crit}, ENT}$ for the maximally entangled states (diagonal form of the correlation tensor)of a pair of qudits:
\begin{equation}
\ket{\psi_{\text{max}}^{d}} = \frac{1}{\sqrt{d}} \sum_{i=0}^{d-1} \ket{ii}
\label{eq:MES}
\end{equation}
We present results for arbitrary $d$ and for $d \to \infty$. 

We also analyze all  non-maximally entangled pure initial states of qutrits ($d=3$), parametrized as follows: 
\begin{eqnarray}
\ket{\psi^3(\alpha,\beta)} &=& \cos\alpha \ket{00}+ \sin\alpha\sin\beta \ket{11} \nonumber \\ &+& \sin\alpha\cos\beta \ket{22}.
\label{eq:NMES}
\end{eqnarray}

The study is also done for lower Schmidt rank states, for $d=3, 4$:
\begin{equation}
\ket{\psi^{d}_{\textrm{rank-}k}}= \sum_{j=d-k}^{d-1}c_j \ket{jj},
\label{eq:SR_state}
\end{equation}
where $k<d$ and $\sum_{j=d-k}^{d-1}|c_j|^2 =1.$

\section{ Results for various types of noise}
\subsection{White noise}
\label{wn}

White noise is represented by an admixture of the totally mixed state. A mixture of white noise and two-qudit pure state $\ket{\psi}$ can be written as
\begin{equation}
\rho_{\text{white}}(v)=v |\psi\rangle \langle \psi | +\frac{1-v}{d^2} \openone_{d} \otimes \openone_{d}.
\label{eq:white}
\end{equation}

Since white noise has no correlations, the correlation tensor $\hat{T}(v)$ for the state $\rho_{\text{white}}(v)$ is given by $\hat{T}(v)= v \hat{T}$. 
In the case of the maximally entangled state, only diagonal components of the tensor are nonzero and given by $T_{ii}(v)=\pm v/(d-1)$ for $i\in\{1, 2, \dots, d^2-1\}$. Therefore, we have $L_{\text{max}}=v/(d-1)$ and $||\hat{T}(v)||^2=v^2(d+1)/(d-1)$. Using the criterion~(\ref{eq:ent_criterion0}), we can conclude that the state (\ref{eq:white}) is entangled if $v>(d+1)^{-1}$. This recovers the result of Ref.~\cite{Werner89}. In the limit of $d \to \infty$, the critical value $v^d_{\text{crit}, ENT}$ tends to 0.

\begin{figure}
\centering
\includegraphics[width=7cm]{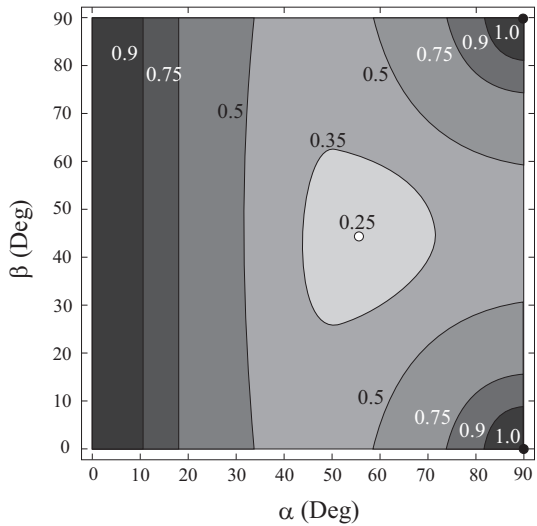}
\caption{The critical values $v_{\text{crit}}$ to detect entanglement for two-qutrit non-maximally entangled states~(\ref{eq:NMES}) and white noise admixture. For $v>v^3_{\text{crit}, ENT}$, the state is entangled. The lowest critical value, $v^3_{\text{crit}, ENT} = 0.25$, is obtained for the maximally entangled state $\ket{\psi^{3}_{\text{max}}}$.}
\label{twoqutrits}
\end{figure}

For the case of  two-qutrit non-maximally entangled states~(\ref{eq:NMES})we obtain $L_{\text{max}}=v(8+\sqrt{K})/16$ and $||\hat{T}(v)||^2=v^2(2-K/64)$, where $K=31+12\cos2\alpha+21\cos4\alpha+24\cos 4\beta\sin^4\alpha$. In Fig.~\ref{twoqutrits}, we present the critical values $v^3_{\text{crit}, ENT}$. The pure state (\ref{eq:NMES}) is entangled for all $\alpha$ and $\beta$ except for $\alpha=0$, $\{\alpha=\pi/2, \beta=0\}$, and $\{\alpha=\pi/2, \beta=\pi/2\}$. The lowest value of $v^3_{\text{crit}, ENT} = 0.25$ is obtained by the maximally entangled state $\ket{\psi_{\text{max}}^{3}}$.

In Table~\ref{chart1} we  present the numerical results  for states $\eqref{eq:SR_state}.$  They are for  two-qutrit states of the Schmidt  form $\frac{1}{\sqrt{2}}(|11\rangle+|22\rangle),$ and of  two-ququart states $\frac{1}{\sqrt{2}}(|22\rangle+|33\rangle)$ and $\frac{1}{\sqrt{3}}(|11\rangle+|22\rangle+|33\rangle).$ All this is compared with proper maximally entangled states. The analysis shows that the lower is the Schmidt rank of such states the more fragile is their entanglement.

%--------------------------
\subsubsection{White noise treated as local depolarizing noise }
%--------------------------
A $d$-dimensional quantum state exiting  a local depolarizing channel is given by
\begin{eqnarray}
\Lambda_{r}^{\text{depol}}(\rho)=(1-r)\rho +\frac{r}{d} \openone.
\end{eqnarray}
 The use of such a formalism gives us a method of estimating how much noise (disturbance) is introduced per qudit. With probability $1-r$ the initial state $\rho$ still remain unaltered by the decoherence. To describe the local depolarizing effect on a two-qudit system, one can use the following  transformation of the generalized Gell-Mann matrices:
\begin{eqnarray}
\label{Gelldepol1} 
\Lambda_{r}^{\text{depol}}(M_k)= (1-r) M_k,
\end{eqnarray}
where  $k \in \{1, 2, \dots, d^2-1 \},$ and make the replacement in $T_{ij}= c(d) \textrm{Tr} [\rho (M_{i} \otimes M_{j})].$

For the  maximally entangled state~(\ref{eq:MES}), after such a transformation only diagonal components of the correlation tensor are nonvanishing, and they read
\begin{equation}
\label{corredepol-d1}
T_{ii} (r)=\pm\frac{(1-r)^2}{(d-1)}~~~\text{for}~i \in \{1, 2, \dots, d^2 -1\}. 
\end{equation}
Therefore, we have $L_{\text{max}}=(1-r)^2/(d-1)$ and $||\hat{T}(r)||^2=(1-r)^4(d+1)/(d-1)$. We define  $1-r=p$, the critical value $p^d_{\text{crit}, ENT}$ in local depolarized channel is obtained as
\begin{equation}
\label{depol-v}
p^d_{\text{crit}, ENT}=(d+1)^{-\frac{1}{2}}.
\end{equation}
We see that when $d\to\infty$, $p^d_{\text{crit}, ENT} \to 0$. Thus with increasing {\em d} the maximal entanglement is more and more resistant with respect to the depolarizing noise.  

\begin{figure}
\centering
\includegraphics[width=7cm]{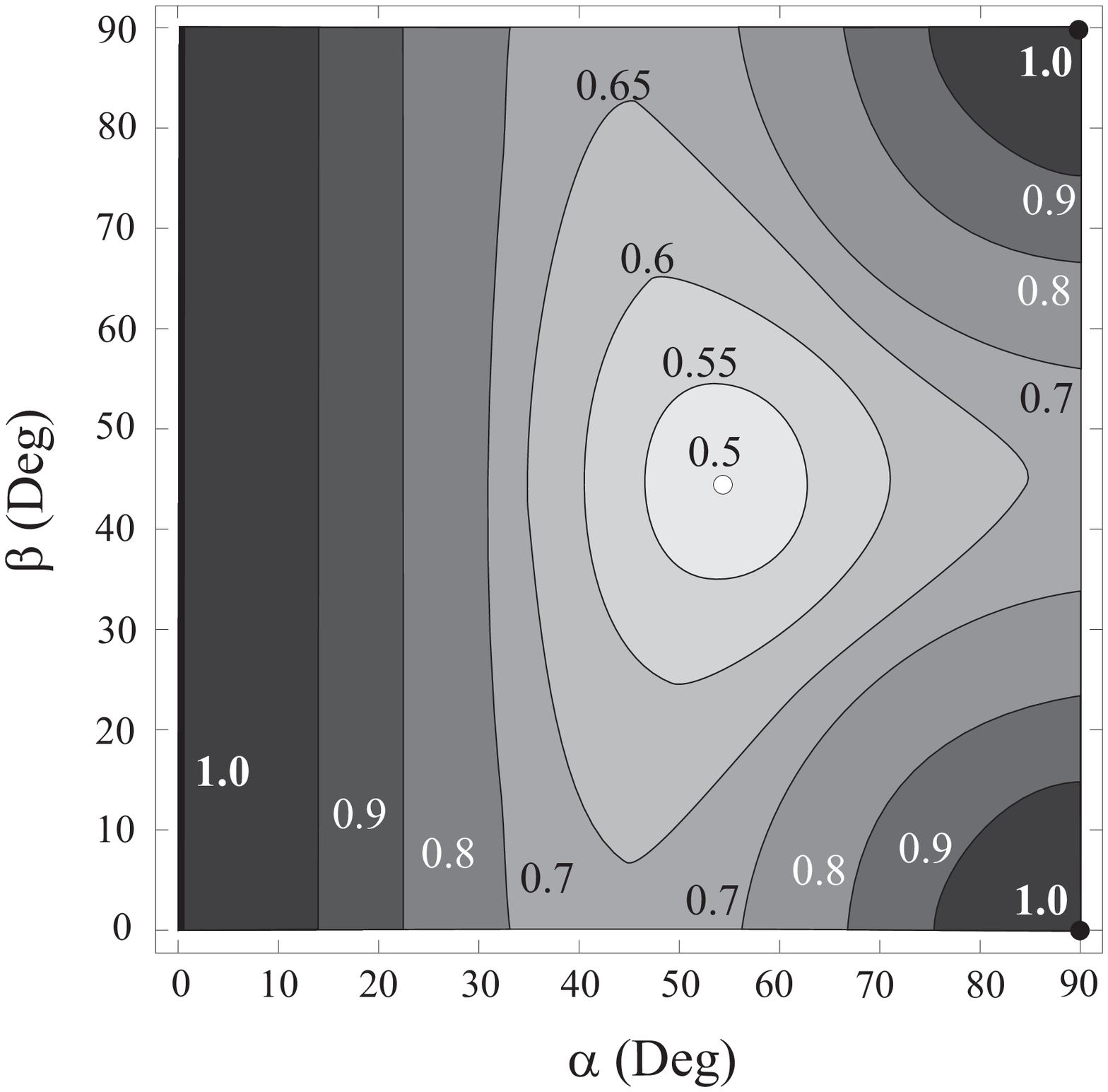}
\caption{Critical values $p^3_{\text{crit}, ENT}$ for entanglement detection are shown for two-qutrit non-maximally entangled state under local depolarizing noise.  The lowest critical value, $p^3_{\text{crit}, ENT} = 0.5$, is achieved for the maximally entangled state $\ket{\psi^{3}_{\text{max}}}$.}
\label{twoqutritsdepol}
\end{figure}

Our study for arbitrary (initially) pure states of two qutrits (\ref{eq:NMES}) gives   $L_{\text{max}}=p^2(8+\sqrt{K})/16$ and $||\hat{T}(p)||^2 =p^4(2-K/64)$. The critical values $p^d_{\text{crit}, ENT}$ are ploted in Fig. \ref{twoqutritsdepol}. It is entangled for all $\alpha$ and $\beta$ except those related to product states: $\alpha=0$, $\{\alpha=\frac{\pi}{2}, \beta=0\}$ and $ \{\alpha=\frac{\pi}{2}, \beta=\frac{\pi}{2}\}$. The lowest value of $p^3_{\text{crit}, ENT} = 0.5$ is obtained by the maximally entangled state~(\ref{eq:MES}). 
In Tab.~\ref{chart} we give numerical results of the critical values $p^d_{\text{crit}, ENT}$ for lower Schmidt rank states  $\eqref{eq:SR_state}$ for $d=3, 4.$ We see that for a specific dimension,  entangled states of  lower Schmidt rank  are less resistant to local depolarizing noise  than  entangled states  of higher Schmidt rank. Thus the new analysis does not change the picture much.

 For local depolarizing channel $||\hat{T}(p^d_{\text{crit}, ENT})||^2=\frac{1}{d^2-1}$ and $||\hat{T}(p=1)||^2=\frac{d+1}{d-1}$ for the maximally entangled state  which yields $\xi(p^d_{\text{crit}, ENT})=\frac{1}{d+1}.$ This is identical to $v^d_{\text{crit}, ENT}$ to detect entanglement  for the maximally entangled state in a global depolarizing noisy channel (white noise).

%--------------------------- 
\subsection{Product noise}
\begin{figure}
\centering
\includegraphics[width=7cm]{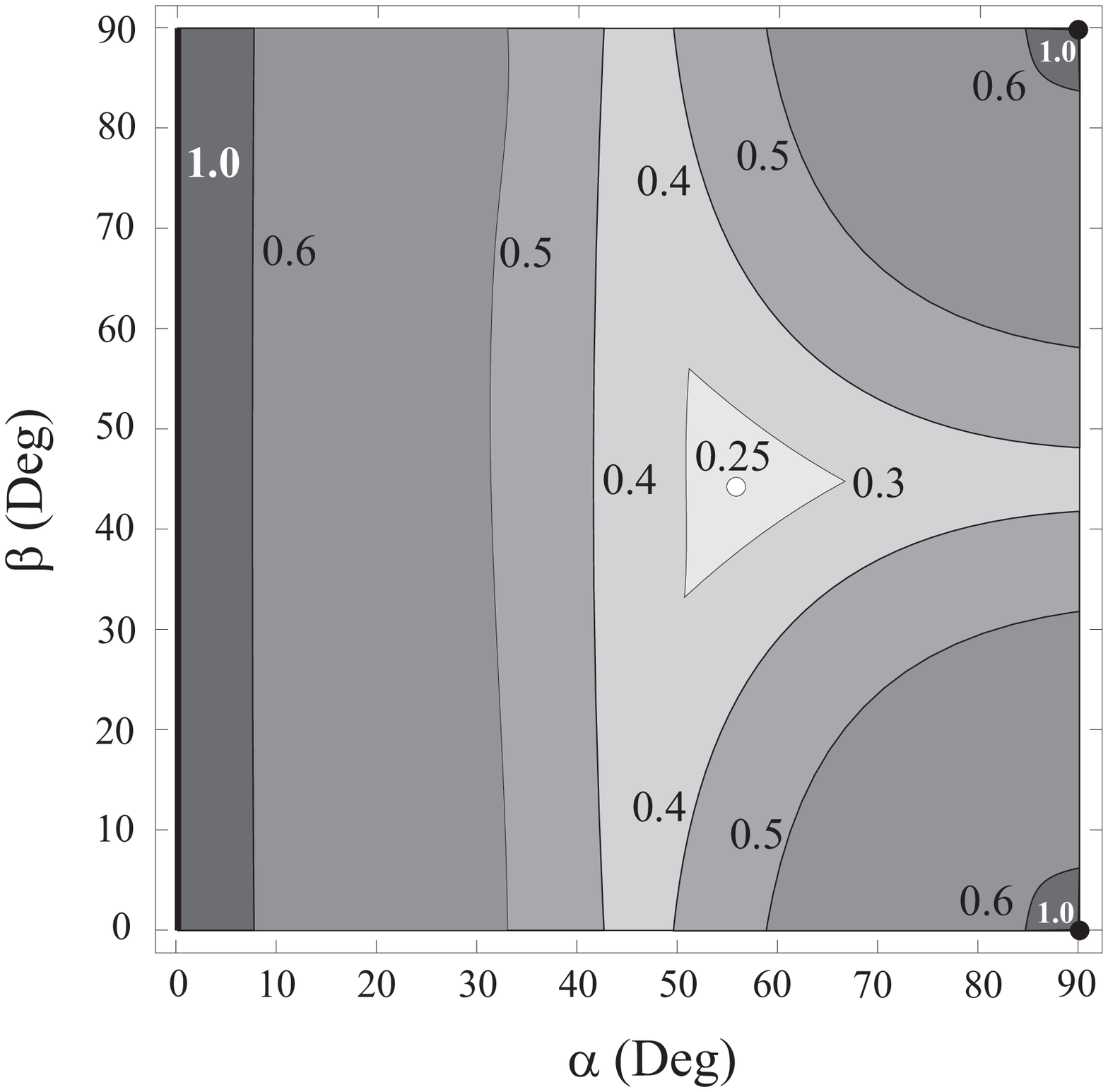}
\caption{The critical values $v^3_{\text{crit}, ENT}$ to indicate entanglement for product noise. For $\alpha \to 0$, $\{\alpha \to \pi/2, \beta \to 0 \}$, and $\{\alpha \to \pi/2, \beta \to \pi/2\}$, the state $\eqref{prod}$ is entangled for the critical visibility approaching $0.6039$.}
\label{twoqutritsprod}
\end{figure}

%---------------------------
We consider the following output states:
\begin{equation}
\label{prod}
\rho_{\text{prod}} (v)=v \ket{\psi} \bra{\psi}+(1-v)\rho_a\otimes\rho_b,
\end{equation} 
where $\rho_i= \sum_{j=0}^{d-1}|c_j|^2 |j\rangle_i\langle j|$ is the reduced density matrix of the $i$th subsystem. The critical values $v^3_{\text{crit}, ENT}$ are presented in Fig.~\ref{twoqutritsprod}. The lowest critical value $v^3_{\text{crit}, ENT}=0.25$ is also obtained by the maximally entangled state $\ket{\psi^{3}_{\text{max}}}$ (for this  state  the product noise is reduced to white noise)
Note that $v^3_{\text{crit}, ENT}=1$ for product states, whereas in the limits $\alpha \to 0$, $\{\alpha \to \pi/2, \beta \to 0 \}$, or $\{\alpha \to \pi/2, \beta \to \pi/2\}$, the state (\ref{prod})  is entangled for the critical visibility approaching 0.6039. In other words, our geometrical method is very sensitive  in the  presence of product noise.

%---------------------------
\subsection{Colored product noise}
%--------------------------
In the special case of output states
\begin{eqnarray}
\label{cn}
\rho_{\text{c}}(v) = v\ket{\psi_{\text{max}}^{d}}\bra{\psi_{\text{max}}^{d}} + (1-v) \hat{\pi}_c\otimes \hat{\pi}_c,
\end{eqnarray}
with  $\hat{\pi}_c=\ket{d-1}\bra{d-1},$ the correlation tensor of the state ($\ref{cn}$) has the following non-vanishing components:
\begin{equation}
T_{ii}(v)=\left\{
\begin{array}{ll}
  \pm \,\frac{v}{d-1} \,\,\,\,\,\, &\text{for} \,\,\, i \in \{1,\dots, d^2 -2\}, \\
  1-\frac{v(d-2)}{d-1} \,\,\,\,\,\, &\text{for} \,\,\, i = d^2 -1.
\end{array}\right.
\end{equation}
In order to get more efficient entanglement condition,  we consider a generalized version of the criterion of Eq.~($\ref{eq:ent_criterion0}$)
\begin{eqnarray}
\max_{\hat{T}^{\text{prod}}} (\hat{T}, \hat{T}^{\text{prod}})_G < ||\hat{T}||_{G}^{2}, 
\label{eq:ent_criterion1_modi}
\end{eqnarray}
where now the scalar product is defined by  a positive semidefinite diagonal metric $G$~\cite{Badziag08}:
\begin{equation}
\label{scalar1}
(\hat{X}, \hat{Y})_{G} = \sum_{i, j=1}^{d^2-1} X_{ij} G_{ij} Y_{ij}.
\end{equation}
We consider a diagonal metric $G$ with the following nonzero elements: $G_{ii} = 1$ for $i \in \{1, 2, \dots, d^2 -2\}$ and $G_{d^2 -1 d^2 -1} = v$. With such a metric, the left-hand side ($\ref{eq:ent_criterion1_modi}$) reads $L_{\text{max}}^{G}=v[1-v(d-2)/(d-1)]$ and $||\hat{T}(v)||_{G}^{2}$ as $v[1+v(-6+6d-d^2 + v(d-2)^2)/(d-1)^2]$. For $d \geq 2$, $||\hat{T}(v)||_{G}^{2}$ is always greater than $L_{\text{max}}^{G}$. It implies that the state $\rho_{\text{c}}(v)$ is always entangled except $v=0$, i.e., even for an infinitesimally small $v$ the state is entangled.

%--------------------------
\subsection{Amplitude damping noise}
\label{sec:AD}
%--------------------------
An amplitude damping channel can be described by a process of  energy dissipation of a quantum system to environment. The transition of excitation occurs between excited state  and the ground state with a finite probability. No transition  is allowed between excited states. Such an amplitude damping channel can be described by:
\begin{eqnarray}
\rho_{\text{AD}}(r)=\!\!\!\!\sum_{k, m=0}^{d-1}{E_k(r)\otimes E_m(r)\rho( E_k (r)\otimes E_m(r))^\dagger},
\label{trans}
\end{eqnarray}
where
\begin{eqnarray}
\label{kamp}
E_0 &=& \ket{0}\bra{0} + \sqrt{1-r}\sum_{i=1}^{d-1} \ket{i}\bra{i}, \nonumber\\
E_{j}&=&\sqrt{r}~\ket{0}\bra{j},~~~ j \in \{1,2,\dots,d-1\}.
\end{eqnarray}
The transformation of the generalized Gell-Mann matrices is given by
\begin{eqnarray}
M_{j} (r) = \left \{ \begin{array}{ll}
\sqrt{1-r} M_{j}~~&\text{for}~M_j \in \{M_{1,k}^{s(\text{or}\,a)} \}\\
(1-r) M_{j}~&\text{otherwise}, 
\end{array} \right.
\label{eq:gell_mann_trans}
\end{eqnarray}
where $j \in \{1, 2, \dots, d^2 -d \}, k \in \{2, 3, \dots, d\}$,  the $M_{1,k}^{s(\text{or}\,a)}$ is the symmetric (or antisymmetric) Gell-Mann matrices (see Appendix~\ref{apx:GM_matrix} for details). Here,  we do not consider the transformation rule for diagonal Gell-Mann matrices as they are redundant in  later calculation.

For the two-qudit maximally entangled state (\ref{eq:MES}), we have the following  nonzero components of the correlation tensor
\begin{eqnarray}
T_{ii} (r)=
\begin{cases}
\pm\frac{(1-r)}{(d-1)}~~&\text{for}~M_i \in \{M_{1,k}^{s(\text{or}\,a)} \} \\
\pm\frac{(1-r)^2}{(d-1)}~~ &\text{otherwise},
\end{cases}
\label{eq:corre_trans_AD}
\end{eqnarray}
where $i \in \{1, 2, \dots, d^2 -d \}$. Although $T_{ij} (r)$ are non zero for $i, j \in \{ d^2 -d,  d^2 -d+1, ...d^2 -1 \},$ still they do not have any impact in our calculation to obtain the  {\em sufficient } condition for entanglement. More precisely,  to calculate the critical value $p_{\text{crit}}$ for amplitude damping noise with a {\em specific} combination of components of  correlation tensor, we use the generalized criterion~(\ref{eq:ent_criterion1_modi}) involving the diagonal metric; $G_{ii} = 1$ for $i \in \{1, 2, \dots, d^2 -d\}$ and other components are zero.  As a result, we have $L_{\text{max}}=(1-r)/(d-1)$ and $||\hat{T}(r)||^2=(1-r)^2 \left[2+(d-2)(1-r)^2 \right]/(d-1)$.
%Thus, effectively $|\pm(1-r)/(d-1)|$ appears $2(d-1)$ times  whereas $|\pm(1-r)^2/(d-1)|$ for $(d-1)(d-2)$ times. Because  $0\leq r \leq 1$, 
Thus, the entanglement criterion (\ref{eq:ent_criterion1_modi}) reduces to $(d-2)p^3+2p>1$, where $p=1-r$. We find that for $d>2$ the state is entangled if the value $p$ exceeds the critical value
\begin{equation}
p^d_{\text{crit}, ENT}=\left( \frac{1}{2(d-2)} \right)^{1/3}  \left( 1-\frac{A^{1/3}}{B^2} \right) B,
\label{eq:crit_AD}
\end{equation}
where $A = 2^5/[3^3 (d-2)]$ and $B = [1 + (1+A)^{1/2}]^{1/3}$. The critical value $p^d_{\text{crit}, ENT}$ goes to zero when $d \to \infty$.

\begin{figure}
\centering
\includegraphics[width=7cm]{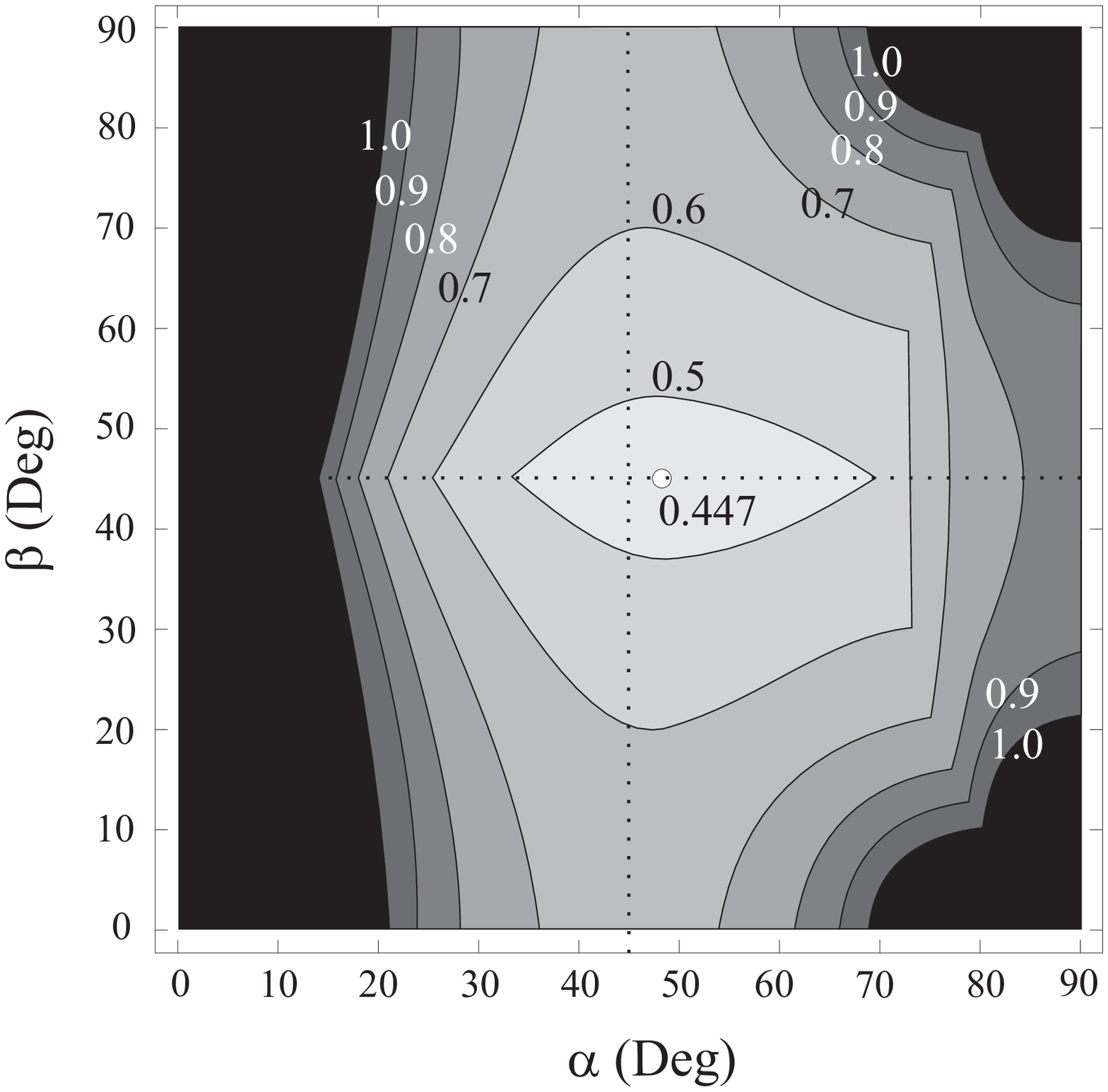}
\caption{We show the critical value $p^3_{\text{crit}, ENT}$ of entanglement identification for non-maximally entangled state of qutrits for amplitude damping noise. The lowest value( 0.447(approximate)) for $p^3_{\text{crit}, ENT}$ is obtained for the state $\ket{\psi(\frac{4\pi}{15},\frac{\pi}{4})}$(\ref{eq:NMES}) }
\label{twoqutritsad}
\end{figure}

It is worth noting that non-maximally entangled states for $d = 3, 4$ in Eq.~(\ref{eq:NMES}) are more robust than the maximally entangled state (\ref{eq:MES}) against this type of noise. The corresponding non-maximally entangled states are $\ket{\psi(\frac{4\pi}{15},\frac{\pi}{4})}$ for $d=3$ and $\ket{\psi^4_{\text{nmax}}}=\cos (0.853) \ket{00}+\frac{1}{\sqrt{3}} \sin (0.853) ( \ket{11}+\ket{22}+ \ket{33}).$ The critical values $p^3_{\text{crit}, ENT}$ are ploted in Fig. \ref{twoqutritsad}.  In Table~\ref{chart} we show numerical results of  the critical values $p^d_{\text{crit}, ENT}$ for lower Schmidt rank states $\eqref{eq:SR_state}$ for $d=3, 4.$ It is found  that the maximally entangled states~ $\eqref{eq:MES}$ of lower Schmidt rank  are less robust than the maximally entangled states of higher Schmidt rank. As a matter of fact the positive semidefinite diagonal metric $G$ is not unique, one can obtain lower values for $p^d_{\text{crit}, ENT}$ than the results presented in our manuscript with the optimal  $G.$

\begin{figure}
\centering
\includegraphics[width=7cm]{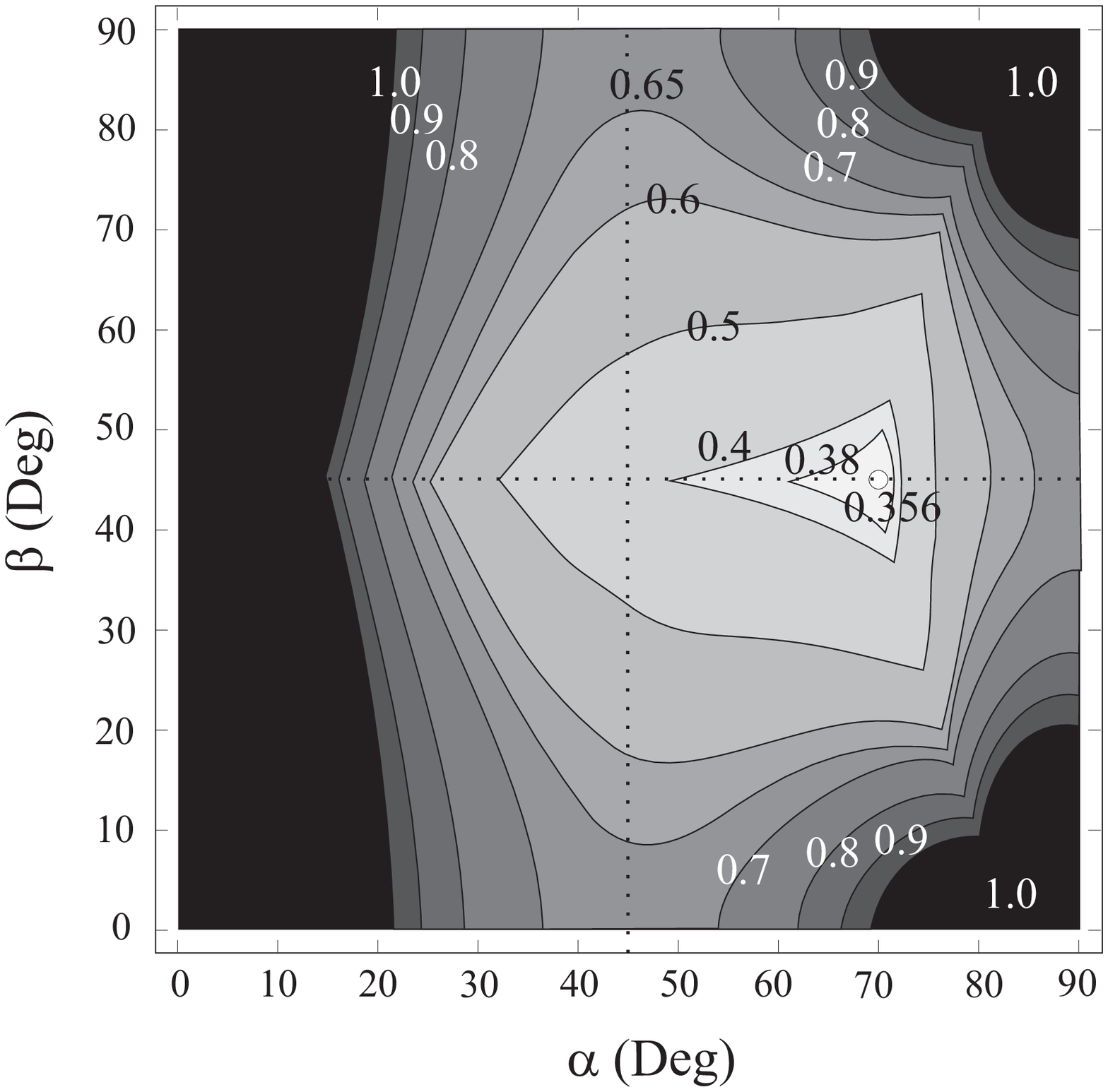}
\caption{We show the critical value $\xi(p^3_{\text{crit}, ENT})$ of entanglement identification for non-maximally entangled state of qutrits for amplitude damping noise.  The lowest value (0.3560) for $p^3_{\text{crit}, ENT}$ is obtained for the state $\ket{\psi(\frac{7\pi}{18},\frac{\pi}{4})}$(\ref{eq:NMES}) }
\label{twoqutritsad1}
\end{figure}

Also, we check critical parameter in $\eqref{square}$ for the maximally entangled state $\eqref{eq:MES}.$ Here $||\hat{T}(p^d_{\text{crit}, ENT})||^2=\frac{(p^d_{\text{crit}, ENT})^2}{d-1}(2+(d-2)(p^d_{\text{crit}, ENT})^2)$ and $||\hat{T}(p=1)||^2=\frac{d}{d-1}$  which yields $\xi(p^d_{\text{crit}, ENT})=p^d_{\text{crit}, ENT}(2+(d-2)(p^d_{\text{crit}, ENT})^2)^{\frac{1}{2}}d^{-\frac{1}{2}}.$  Hence, $\xi(p^d_{\text{crit}, ENT})\to 0$, when $d \to \infty.$ Due to  our entanglement criterion  we obtain lower value  for $\xi(p^3_{\text{crit}, ENT})$ for  a specific non-maximally entangled state Eq.~(\ref{eq:NMES}) than the maximally entangled state (\ref{eq:MES}) while analyzing for amplitude damping noise. The corresponding non-maximally entangled state is $\ket{\psi(\frac{7\pi}{18},\frac{\pi}{4})}$ for $d=3.$  For this specific state $\ket{\psi(\frac{7\pi}{18},\frac{\pi}{4})},$ we obtain $p^3_{\text{crit}, ENT}=0.4990.$ The critical values $\xi(p^3_{\text{crit}, ENT})$  are plotted in Fig. \ref{twoqutritsad1}. 
A comparison between  values of $v^d_{\text{crit}, ENT}$ and $ \xi(p^d_{\text{crit}, ENT})$ parameter, calculated for white or local depolarizing noise and amplitude damping noise,  respectively is shown in  Table~\ref{chart1} .

\begin{table}
\centering 
\begin{tabular}{l|c|ccc|c}
\hline\hline
noise & ~$d$~ & $\ket{\psi_{\text{max}}^{d}}$ & $\ket{\psi^{d}_{\text{rank-}2}}$ & $\ket{\psi^{d}_{\textrm{rank-}3}}$ & $\ket{\psi^d_{\text{nmax}}}$\\

\hline
                     & ~2~ &  0.5773   & -          & -             &-     \\
local depolarizing & ~3~ &  0.5         & 0.6546 &  -            &-     \\
                     & ~4~ &  0.4472   & 0.6794 & 0.5283   & -   \\
\hline
				& ~2~  &  0.5	   & -		 & - 		& -		  \\
amplitude damping   & ~3~  &  0.4534 & 0.8165 & -	         & 0.4465	  \\
				& ~4~  &  0.4239 & 0.8660 & 0.6124 & 0.4074     \\
\hline\hline
\end{tabular}
\caption{Summary of the critical values $p^d_{\text{crit}, ENT}$ to detect entanglement for the various bipartite states under  depolarizing and amplitude damping noises: the maximally entangled state $\ket{\psi_{\text{max}}^{d}}$, the non-maximally entangled state $\ket{\psi^d_{\text{nmax}}}$, and the lower Schmidt rank states $\ket{\psi^{d}_{\text{rank-}k}}$. For amplitude damping noise, the specific non-maximally entangled states present lower  critical value $p^d_{\text{crit}, ENT}$ than the maximally entangled ones. The specific states are presented below Eq.~(\ref{eq:crit_AD}).}
\label{chart}
\end {table}

\begin{table}
\centering 
\begin{tabular}{l|c|ccc|c}
\hline\hline
noise & ~$d$~ & $\ket{\psi_{\text{max}}^{d}}$ & $\ket{\psi^{d}_{\text{rank-}2}}$ & $\ket{\psi^{d}_{\textrm{rank-}3}}$ & $\ket{\psi^d_{\text{nmax}}}$\\

\hline
                     & ~2~ &  0.3333   & -          & -             &-     \\
 local depolarizing & ~3~ &  0.25         & 0.4285 &  -       &-     \\
                     & ~4~ &  0.2   & 0.4615 & 0.2790     \\
\hline
				& ~2~  &  0.5	   & -		 & - 		   &-     \\
amplitude damping   & ~3~  &  0.3888 & 0.6667 & -	       	 & 0.3560	  \\
				& ~4~  &  0.3256 & 0.750 & 0.3750    & -		  \\
\hline\hline
\end{tabular}
\caption{A summary of the critical values $ \xi(p^d_{\text{crit}, ENT})$ for  local depolarizing noises, and   amplitude damping noise in  detection of  entanglement for  $\ket{\psi_{\text{max}}^{d}}$  and the lower Schmidt rank states $\ket{\psi^{d}_{\text{rank-}k}}$  is given in a tabular form. As a mater of fact,  $ \xi(p^d_{\text{crit}, ENT})$ computed for each cases with local depolarizing noise  is identical to $ v^d_{\text{crit}, ENT}$ obtained for the identical states  with white noise.  Also,   for amplitude damping noise, the specific non-maximally entangled state presents  lower  critical values $\xi(p^3_{\text{crit}, ENT})$ than the maximally entangled one. The specific state is $\ket{\psi(\frac{7\pi}{18},\frac{\pi}{4})}$ for $d=3.$}
\label{chart1}
\end {table}

%------------------------------------------------
\section{Testing Entanglement with Bell inequalities}
%------------------------------------------------
In this section, we analyze noise resistance  of violation of  Bell-type inequalities for the bipartite quantum states (\ref{state0}).  To this end, we employ the CGLMP inequality introduced in Ref.~{\cite{CGLMP}}, which can be written as
\begin{eqnarray}
\label{Bell}
I_d= \sum_{k=0}^{\lfloor\frac{d}{2}\rfloor-1}\left(1-\frac{2k}{d-1}\right) \left(\mathcal{B}_{k} -\mathcal{B}_{-(k+1)}\right) \leq 2,
\end{eqnarray}
where $\mathcal{B}_{k} = P(A_1=B_1+k) + P(B_1=A_2+k+1)
+P(A_2=B_2+k) +P(B_2=A_1+k)$.

We compare  robustness of  violation  of the CGLMP inequality, with respect to  white noise,   in case of the maximally entangled state $\ket{\psi^d_{\text{max}}}$  (\ref{eq:MES}),  presented   in \cite{CGLMP} with  robustness of  violation by  the same state under different noisy channels. 

%------------------------------------------------
\subsection{White noise}
%------------------------------------------------
In this section we give an analytical solution for  the critical visibility to violate the inequality  $\ref{Bell}$ with the maximally entangled states and then consider the case  for $d\to \infty.$ 

Since white noise does not contribute to the violation of the CGLMP inequality~(\ref{Bell}) as stated  in Ref.~(\ref{Bell}), the quantum value for the state $\rho_{\text{white}} (v)$ in Eq.~(\ref{eq:white}) is given by 
\begin{equation}
I^d_{QM}[\rho_{\text{white}} (v)] = v I^d_{QM}[|\psi\rangle].
\end{equation}
Therefore the critical parameters for the violation of the CGLMP inequality~(\ref{Bell}) are given by
\begin{equation}
\label{wnn}
v_{crit, LR}^d = 2/ I^d_{QM}[|\psi\rangle].
\end{equation}
and  equal: $v_{crit, LR}^2\approx 0.7073$, $v_{crit, LR}^3 \approx0.6962$, $v_{crit, LR}^4 \approx 0.6906$ and $ v_{crit, LR}^{\infty}\approx 0.6734$ \cite{CGLMP}. Note that the lowest critical visibility for the CGLMP inequality is achieved by non-maximally entangled states~\cite{ACIN} and studied in details (for $d=3$) in Ref.~\cite{GRUCA2}.  The lowest critical visibility ($v^3_{\text{crit}, LR}$=0.6861)(Ref.~\cite{GRUCA2}) to violate CGLMP inequality in $d=3$ is obtained for non-maximally entangled state $\ket{\psi^3(1.0601, 0. 7854)}$ (see ~$\eqref{eq:NMES}).$ For this specific non-maximally entangled state $\ket{\psi^3(1.0601, 0. 7854)},$ we obtain $ v^3_{\text{crit}, ENT} =0.2883.$
%------------------------------------------------
\subsubsection{White noise treated as local depolarizing channel}
%------------------------------------------------
In this section  we check the robustness  of CGLMP inequality in a local depolarizing channel with maximally entangled state $\ket{\psi^d_{\text{max}}}$.  

Following the transformation rules for the Gell-Mann matrices shown in $\ref{Gelldepol1}$, the state $\ket{\psi^d_{\text{max}}}$ after local depolarizing channel can be written as
\begin{equation}
\rho_{\text{depol}}(r)=(1-r)^2\ket{\psi^d_{\text{max}}}\bra{\psi^d_{\text{max}}}+\frac{r(2-r)}{d^2} \openone_{d} \otimes \openone_{d}.
\label{dee}
\end{equation}
 Since the noise part does not contribute to the Bell expression (\ref{Bell}) with quantum mechanical observables,  leads to the following condition for the critical parameters for the violation of (\ref{Bell}):
\begin{equation}
p_{crit, LR}^d = \sqrt{2/ I^d_{QM}[|\psi\rangle]},
\end{equation}
where $p^d_{crit, LR}=(1-r^d_{crit, LR}).$
%where $I_{QM}(d)$ is given in the Appendix $\ref{quantum'}.$ 
The results are given in Tab. \ref{t}. However  in terms of visibility $v^d_{crit, LR}$, the result is identical with $\eqref{wnn}.$

 %For $d$ $\to$  $\infty$, $v_{crit}$=0.82063. Here  $p_{crit}$ is a decreasing function of $d$.
\begin{table}
\begin{tabular}{c|c|c}
\hline\hline
~$d$~&depolarizing noise& amplitude damping noise \\
\hline
~2~ & 0.8410 & 0.7071  \\ 
~3~& 0.8344 & 0.7468  \\
~4~ & 0.8310 & 0.7647  \\
~5~ & 0.8290 & 0.7750  \\
~10~ & 0.8248 & 0.7954  \\
~$\infty$~ &0.8206 & 0.8206  \\
\hline\hline
\end{tabular}
\caption{ The critical parameter to indicate violation of CGLMP inequality, $p^d_{crit, LR}$ are calculated for local depolarizing noise and amplitude damping  noise  for the bipartite maximally entangled state in higher dimensional systems.}
\label{t}
\end{table}
%------------------------------------------------
\subsection{Colored noise}
%------------------------------------------------
In this section  we check the violation of CGLMP inequality by the  maximally entangled state , $\ket{\psi^d_{\text{max}}}$~\ref{state} mixed with colored noise~$\ref{cn}$. Numerically we find that for $d= 3$ instead of choosing settings $(\ref{settings})$ if we choose generalized $SU(3)$ transformation matrix,  then $\eqref{Bell}$ is violated whenever $v>0.$

%------------------------------------------------
\subsection{Amplitude damping channel}
%------------------------------------------------

The same as in the previous cases, we examine the violation of the CGLMP inequality with state $\ket{\psi^d_{\text{max}}}$. 

The state $\ket{\psi^d_{\text{max}}}$ after amplitude damping channel~(\ref{kamp}) is given by
\begin{eqnarray}
&&\rho_{\text{AD}}(r) = \nonumber \\
&&\frac{1+r^2(d-1)}{d}\ket{00} \bra{00}
+\frac{(1-r)^2}{d}\sum_{m, k\neq 0}^{d-1}\ket{mm} \bra{kk} \nonumber \\ 
&+&\frac{(1-r)}{d}\left[\sum_{m\neq 0}^{d-1} \ket{mm} \bra{00}
+\sum_{k\neq 0}^{d-1}\ket{00} \bra{kk}\right] \nonumber \\ 
&+&\frac{r(1-r)}{d}\left[\sum_{m\neq 0}^{d-1}\ket{m0} \bra{m0}
+\sum_{k\neq 0}^{d-1}\ket{0k} \bra{0k}\right].
\label{state1}
\end{eqnarray}

%Following the derivation shown in Appendix~\ref{C}, the maximum value of $I_d$ $\ref{Bell}$ equal to $2$. 

We numerically check the critical parameters for $d=2, 3, 4, 5, 10$. The results are given in Table~\ref{t}.

We see that the critical parameter increases with the dimension. It has a tendency to attain saturation for very large $d$. We give an analytical result for $d$ tends to $\infty$ in Appendix~(\ref{C}). However, the lowest critical visibility ($p^3_{\text{crit}, LR}$=0.7316) to violate CGLMP inequality in $d=3$ is obtained for non-maximally entangled state $\ket{\psi^3(0.8730, 0. 6449)}$ (see ~$\eqref{eq:NMES}).$ For this specific non-maximally entangled state  we obtain $p^3_{\text{crit}, ENT} =0.5,$ whereas $\xi(p^3_{\text{crit}, ENT})=0.4513.$ Also we check $p^3_{\text{crit}, LR}$  for the  specific non-maximally entangled state $\ket{\psi^3(\frac{4\pi}{15}, \frac{\pi}{4})},$ which gives the lowest $p^3_{\text{crit}, ENT}.$  We obtain  $p^3_{\text{crit}, LR}=0.7429$ for the state. We extend our calculation by checking $p^3_{\text{crit}, LR}$ for the non-maximally state $\ket{\psi^3(\frac{7\pi}{18}, \frac{\pi}{4})}$ for which we obtain the lowest $\xi(p^3_{\text{crit}, ENT})$ considering  all two-qutrit states. The value is obtained as $0.8640.$

Although for $d \to \infty$, $p^{\infty}_{\text{crit}, LR}$ to violate $\eqref{Bell}$  in presence of local  depolarizing noise coincides (see Tab.~\ref{t})  with the same parameter obtained  in case of  amplitude damping noise  but in lower dimensions  the variation in $p^{d}_{\text{crit}, LR}$ in case of amplitude damping noise differs with the one for local  depolarizing noise.

\subsection{ Fidelity of quantum states with critical noise}
In this section  we give analytic expression  of fidelity, i.e., the measure of closeness between two quantum states, e.g.,  the state  after noisy channel  with the pure state (here we only consider the maximally entangled state~$\eqref{eq:MES})$ of the system under investigation.

Usually fidelity $F$ for two quantum states  is defined as $\sqrt{\langle \phi| \rho|\phi\rangle},$ where $\phi$ is a pure state whereas $\rho$ is a mixed state. We choose $\eqref{eq:MES}$ as $\phi$, and $\rho$ is parametrized by  critical parameters $v^d_{\text{crit}, LR}$  for white noise and $ r^d_{\text{crit}, LR}$ for local depolarizing and amplitude damping noises. The general form of $\rho$'s are given in $\eqref{eq:white}$, $\eqref{dee}$, and $\eqref{state1}$ for white , local depolarizing,  and amplitude damping channels, respectively. The corresponding  critical fidelity is written as:

\begin{widetext}
\begin{eqnarray}
F^d_{\text{crit}, LR}=
\begin{cases}
(v^d_{\text{crit}, LR} +\frac{1-v^d_{\text{crit}, LR}}{d^2})^{\frac{1}{2}}&\text{for white noise,} \\
((1-r^d_{\text{crit}, LR})^2 +\frac{r^d_{\text{crit}, LR}(2-r^d_{\text{crit}, LR})}{d^2})^{\frac{1}{2}}&\text{for local depolarizing noise}, \\
\frac{1}{d}((1+(d-1)((1-r^d_{\text{crit}, LR})^2 +1)+(d-1)^2(1-r^d_{\text{crit}, LR})^2)^{\frac{1}{2}}&\text{for amplitude damping noise}.
\end{cases}
\label{fidelity}
\end{eqnarray}
\end{widetext}
We obtain identical value  for $F^d_{\text{crit}, LR}$ for white and local depolarizing channel but  it differs with critical fidelity parameter for amplitude damping noise in finite dimension. But, they coincide when $d\to\infty.$ To show comparison between critical values of fidelity parameter for different noises  we write down  our results  for specific dimensions in a tabular form in Tab. \ref{k1}.
\begin{table}
\begin{tabular}{c|c|c}
\hline\hline
$d$ &  local  depolarizing & amplitude damping\\
\hline
2 & 0.8834  & 0.8660   \\
3 &0.8544   & 0.8397   \\
4 & 0.8426  & 0.8298      \\
$\infty$ & 0.8206 & 0.8206     \\
\hline\hline
\end{tabular}
\caption{$F^d_ {crit, LR}$  is  calculated for white,  local depolarizing  and amplitude damping noise for the  bipartite maximally entangled state.}
\label{k1}
\end{table}
\subsection{Werner gap}
In  Table~\ref{k}, we compare  $(p^d_ {crit, LR}-p^d_ {crit, ENT})$  for local depolarizing  and amplitude damping  channels   in case of  bipartite  maximally entangled states for $d=2, 3, 4, \infty.$  As a matter of fact  $(p^d_ {crit, LR}-p^d_ {crit, ENT})$  has  same physical interpretation as ``Werner-gap", which is defined as $v^d_ {crit, LR}-v^d_ {crit, ENT}.$
\begin{table}[h]
\begin{center}
\begin{tabular}{c|c|c}
\hline\hline
$d$ & local  depolarizing & amplitude damping\\
\hline
2 & 0.2637  & 0.2071   \\
3 &0.3344   & 0.2934     \\
4 & 0.3838  & 0.3408      \\
$\infty$ & 0.8206 & 0.8206     \\
\hline\hline
\end{tabular}
\caption{$(p^d_ {crit, LR}-p^d_ {crit, ENT})$  is  calculated for local depolarizing noise and amplitude damping noise for the  bipartite maximally entangled state.}
\label{k}
\end{center}
\end{table}
It is apparent from the chart that,  as we increase the dimension of the system,  the difference  between $p^d_ {crit, LR},$ and $p^d_ {crit, ENT}$ increases. We find that, when $d\to \infty,$   $p^d_ {crit, ENT}\to 0$ for both noisy channels, whereas  $p^d_ {crit, LR}$ increases  with the increase of  the dimension of the system for amplitude damping noise and decreases with the increase of the dimension  for depolarizing noise. The overall effect reflects the fact that,  if violation of  CGLMP  inequality is considered as an indicator to test the entanglement of the {\em two-qudit maximally entangled} state, then the maximally entangled state  violates the inequality for very high value of $p^d_ {crit, LR}$ in comparison to $p^d_ {crit, ENT}.$ The discrepancy between $p^d_ {crit, LR},$ and $p^d_ {crit, ENT}$ increases with the dimension. Therefore, the result indicates  the known fact that not all states which are entangled violate  specific Bell inequalities (in this case CGLMP ones).   

\section{Final remarks}
We analyze entanglement of bi-partite qudit systems using a special class of entanglement identifier defined with correlation tensors $\eqref{eq:ent_criterion0}.$  To employ the criterion,  in some cases we use operator-sum representations  of the transformed states  to describe the action of different  noisy channels. We characterize the entanglement with a set of  parameters, e.g., $ v^d_ {crit, ENT},$ $p^d_ {crit, ENT},$  and $\xi(p^d_ {crit, ENT}).$ All these parameters indicate identical behavior of entanglement  when $d\to\infty.$ On the other hand we investigate violation of  CGLMP inequality for the  specific states and also compare $p^d_ {crit, LR},$ and $p^d_ {crit, ENT}.$ In addition,  we compute fidelity ($F^d_{\text{crit}, LR}$) of states  as a tool to quantify violation of local realism. At the end,  it is worth to make few comments on  the entanglement criterion with amplitude damping noise. May be due to the  specific type of entanglement indicator,  a specific  non-maximally entangled state for $d=3$ yields  lower $\xi({p^d_ {crit, ENT}}),$ than the maximally entangled state of the respective dimension. It will also be  interesting to investigate if it can  be possible to detect entanglement for lower values of $\xi({p^d_ {crit, ENT}}),$ using either $\eqref{eq:ent_criterion0}$ or $\eqref{eq:ent_criterion1_modi}$ for  different types of  amplitude damping channels with different types of Kraus operator representations other than the one used here.

\begin{acknowledgments} 
The work is subsidized from funds for science for years 2012-2015 approved for international co-financed project BRISQ2 by Polish Ministry of Science and Higher Education (MNiSW). AD was initially supported within the International PhD Project ``Physics of future quantum-based information technologies'': grant MPD/2009-3/4 of Foundation for Polish Science. MZ and JR are supported by TEAM project of FNP. AD and JR acknowledge support of BRISQ2 VII FP EU project. WL is supported by NCN Grant No. 2012/05/E/ST2/02352. 
\end{acknowledgments}

\appendix
%--------------------------
\section{Generalized Gell-Mann matrices}
\label{apx:GM_matrix}
%--------------------------
The generalized Gell-Mann matrices in an arbitrary dimension $d$ are the  generators of  the Lie algebra associated to the special unitary group SU($d$). A recipe for the  construction of  the generalized Gell-Mann matrices is given in the next paragraph. 

Let $\lambda_{j, k}$ denote a matrix with a $1$  in the $(j, k)$-th entry and 0 elsewhere. This leads one to define three groups of matrices: (a) symmetric group; $M_{j, k}^s=\lambda_{j,k}+\lambda_{k,j}$ for $1\leq j<k\leq d$. (b) antisymmetric group; $M^a_{j, k}=-i(\lambda_{j, k}-\lambda_{k, j})$ for $1\leq j<k\leq d$. (c) diagonal group; $M_{l}^{d}=\sqrt{\frac{2}{l(l+1)}}(\sum_{j=1}^l\lambda_{j, j}-l\lambda_{l+1, l+1})$ for $1\leq l \leq d-1$. The elements of three groups compose the generators of SU($d$) group as $\mathcal{M} = \{ M_{1,2}^{s}, M_{1,3}^{s}, \dots, M_{1,2}^{a}, M_{1,3}^{a}, \dots, M_{1}^{d}, \dots, M_{d-1}^{d} \}$. For an arbitrary dimension $d$, the total number of Gell-Mann matrices is $d^{2}-1$. For simplicity, we re-index the elements of the above set as $\mathcal{M} = \{M_1, M_2, \dots, M_{d^2 -1} \}$. The components satisfy the relations of tracelessness as $\text{Tr}(M_j) =0$ and orthogonality as $\text{Tr}(M_i M_j)= 2\delta_{ij}$. For example, for $d=3$ the set $\mathcal{M}$ is given by
\begin{eqnarray}
\label{GELL}
&M_1=
\begin{pmatrix}
0 & 1 & 0 \\
1 & 0 & 0\\
0 & 0 & 0
\end{pmatrix},~
M_2=
\begin{pmatrix}
0 & 0 & 1 \\
0& 0 & 0\\
1 & 0 & 0
\end{pmatrix},~\nonumber\\
&M_3=
\begin{pmatrix}
0 & 0 & 0 \\
0 & 0 & 1\\
0 & 1 & 0
\end{pmatrix},~
M_4=
\begin{pmatrix}
0 & -i & 0 \\
i & 0 & 0\\
0 & 0 & 0
\end{pmatrix},~\nonumber\\
&M_5=
\begin{pmatrix}
0 & 0 & -i \\
0 & 0 & 0\\
i & 0 & 0
\end{pmatrix},
M_6=
\begin{pmatrix}
0 & 0 & 0 \\
0 & 0 & -i\\
0 & i & 0
\end{pmatrix},~\nonumber\\
&M_7=
\begin{pmatrix}
1 & 0 & 0 \\
0 & -1 & 0\\
0 & 0 & 0
\end{pmatrix},~
M_8=
\begin{pmatrix}
\frac{1}{\sqrt{3}} & 0 & 0 \\
0 & \frac{1}{\sqrt{3}} & 0\\
0 & 0 &-\frac{2}{\sqrt{3}}
\end{pmatrix},
\end{eqnarray} 
where the first three matrices, $M_1, M_2$ and $ M_3$, are symmetric matrices, $M_4, M_5$ and $ M_6$ are antisymmetric matrices, and $M_7$ and $M_8$ are diagonal matrices.

%--------------------------
\section{Calculation of coefficient $c(d)$}
%--------------------------
\label{B}
A single qudit density matrix is characterized  by $(d^2-1)$ dimensional real vector as
\begin{equation}
\rho=\frac{1}{d}\left(\openone_d + k\sum_{i=1}^{d^2-1}{ \rho_i M_i}\right),
\end{equation}
where $\rho_i = \text{Tr}(\rho M_i)$ and we call $\vec{\rho}=(\rho_1, \dots, \rho_{d^2 -1})$ a Generalized Bloch vector, and $k$ is a normalization coefficient that results in $\abs{\vec{\rho}}=1$ for pure state. One can obtain the coefficient $k=\sqrt{d(d-1)/2}$ as follows: for pure state, $\text{Tr} (\rho^2) =1/d+ 2k^2\abs{\vec{\rho}}^2/d^2 =1$. Therefore, the coefficient is given by $k=\sqrt{d(d-1)/2}$.

The element of correlation tensor for single qudit can be written as $T_j=\frac{ d}{2}\sqrt{\frac{2}{d(d-1)}}$Tr[$\rho M_j]=\sqrt{\frac{d}{2(d-1)}} \text{Tr}[\rho M_j]$. Therefore the coefficient $c(d)$ for bipartite state is given by $c(d)= \frac{d}{2(d-1)}$.

%--------------------------
\section{Correlation tensor for arbitrary bipartite state}
\label{A}
%--------------------------
Here, we  present a detailed form of elements of the correlation tensor for two-qudit states: $\ket{\psi}=\sum_{n=0}^{d-1} \alpha_{n} \ket{nn}$. As previously  mentioned in Sec.~\ref{ct}, the element of correlation tensor is given by $T_{ij}= c(d) \textrm{Tr} [\rho (M_{i} \otimes M_{j})]$, where $M_{i}$ is the generalized Gell-Mann matrix presented in Appendix~\ref{apx:GM_matrix}. Here we do not consider the case where the  elements of correlation tensor denote single particle correlation. We can divide the correlation tensor into two parts as follows;
\begin{equation}
\hat{T}=c(d)\left(
\begin{array}{ccc}
\hat{T}^{\text{zone1}} & 0 \\
0 & \hat{T}^{\text{zone2}}
\end{array}
\right),
\end{equation}
where the $\hat{T}^{\text{zone1}}$ is $d(d-1) \times d(d-1)$ matrix and its component is obtained by the Gell-Mann matrices in the symmetric and antisymmetric groups in Appendix~\ref{apx:GM_matrix}. It has only diagonal elements as

\begin{equation}
{T}_{ii}^{\text{zone1}} = \pm2 \alpha_j \alpha_k~~~~\text{for}~i \in \{1,\dots,d(d-1)\}, \nonumber
\end{equation}

where $j,k \in \{0,1,\dots,d-1 \}$ and $j<k.$

On one hand, the $(d-1) \times (d-1)$ matrix $\hat{T}^{\text{zone2}}$ has not only off-diagonal components but also diagonal ones. They are obtained by the Gell-Mann matrices from the  diagonal group. For the sake of simplicity, here, we re-index the elements of the $\hat{T}^{\text{zone2}}$ matrix:
\begin{widetext}
\begin{equation}
\hat{T}^{\text{zone2}}=\left(
\begin{array}{cccccc}
T^{\text{zone2}}_{d^2 - (d-1),d^2 - (d-1)} & T^{\text{zone2}}_{d^2 - (d-1),d^2 - (d-2)} & \cdots & T^{\text{zone2}}_{d^2 - (d-1),d^2 -1} \\
T^{\text{zone2}}_{d^2 - (d-2),d^2 - (d-1)} & \ddots \\
\vdots \\
T^{\text{zone2}}_{d^2 -1,d^2 - (d-1)} & & \cdots & T^{\text{zone2}}_{d^2 -1,d^2 -1}
\end{array}
\right) \to
\left(
\begin{array}{cccccc}
T^{\text{zone2}}_{1,1} & T^{\text{zone2}}_{1,2} & \cdots & T^{\text{zone2}}_{1,d-1} \\
T^{\text{zone2}}_{2,1} & \ddots \\
\vdots \\
T^{\text{zone2}}_{d -1,1} & & \cdots & T^{\text{zone2}}_{d -1,d -1}
\end{array}
\right).
\end{equation}
\end{widetext}
Then, each element is given by
\begin{eqnarray}
& & T^{\text{zone2}}_{ii}=\frac{2}{i(i+1)} \left( \sum_{n=0}^{i-1} \alpha_n^2 + i^2 \alpha_i^2 \right), \nonumber \\
& & T^{\text{zone2}}_{ij}=\frac{2}{\sqrt{i j (i+1)(j+1)}} \left( \sum_{n=0}^{i-1} \alpha_n^2- i \alpha_i^2  \right), \nonumber \\
& & T^{\text{zone2}}_{ji} = T^{\text{zone2}}_{ij},
\end{eqnarray}
where $i,j \in \{1, \dots, d-1 \}$ and $i<j$.

%--------------------------
\section{$I^{AD}_{QM}(d)$ for amplitude damping noise}
%--------------------------
\label{C}

For amplitude damping noise,  we choose the identical  two measurement settings per party used to show violation of  CGLMP inequality~\cite{CGLMP, Durt01}. The measurement operators $A_s$ for Alice and $B_t$ for Bob  have $d$ possible outcomes with the following eigenvectors:
\begin{eqnarray}
\label{settings}
\ket{a}_{A,s}&=&\frac{1}{\sqrt{d}}\sum_{j=0}^{d-1}\omega^{j(a+\alpha_s)}\ket{j}_A,\nonumber \\
\ket{b}_{B,t}&=&\frac{1}{\sqrt{d}}\sum_{j=0}^{d-1}\omega^{j(-b+\beta_t)}\ket{j}_B,
\end{eqnarray}
where $\omega = \exp(2 \pi i/d)$, $a, b \in \{0, 1,  \cdots,  d-1\}$ and $\alpha_1=0, \alpha_2=1/2, \beta_1= 1/4$, and $\beta_2=-1/4$. This can be realized as follows:
each one of the two parties first performs unitary operation on their subsystems with only diagonal entries $\{0, e^{i\frac{2m\pi}{d}},\cdots, e^{i\frac{2m(d-1)\pi}{d}}\}, $ where $m$$\in\{\alpha_s, \beta_t\}.$  Then one of them performs a discrete Fourier transform  $U_{FT}$ ( the matrix element is written as $(U_{FT)_{pq}}=\frac{1}{\sqrt{d}}e^{\frac{2i\pi pq}{d}}$)  and other one applies conjugate  $U^*_{FT}$  and at last they measure in their original bases. Experimentally it can be realized by using  phase shifter and  multi-port  beam splitter  on each side.
Here, we denote $P_{QM}^{AD} (A_s=a, B_t=b)$ as the joint probability that Alice obtain the outcome $a$ from the measurement $A_s$ and Bob obtain the outcome $b$ by measuring  $B_t$ for a given state $\rho_{AD}(r)$ in Eq.~(\ref{state1}). Then, the joint probability reads

\begin{eqnarray}
&&P_{QM}^{AD} (A_s=a, B_t=b)=
\frac{1-(d-1)(r-2)r}{d^3} +\frac{(1-r)}{d^3} \nonumber \\
&& \times \left[(1-r)\sum_{x, y\neq 0}^{d-1}\omega^{(x-y)\gamma} +\frac{2}{d^3}\sum_{p\neq 0}^{d-1}\cos \left(\frac{2\pi p\gamma}{d}\right)\right],
\label{prob1}
\end{eqnarray}
where $\gamma=a-b+\alpha_s+\beta_t$. After summation, Eq.~$\eqref{prob1}$ is given by
\begin{eqnarray}
\label{probability}
&&P_{QM}^{AD} (A_s=a, B_t=b)=\frac{1-(d-1)(r-2)r}{d^3} +\frac{(1-r)}{d^3} \nonumber \\
&&\times\left[(1-r)\frac{\sin^2 \left(\frac{\pi(d-1)\gamma}{d}\right)}{\sin^2\left(\frac{\pi \gamma}{d}\right)}+\frac{\sin\left(\frac{\pi(2d-1)\gamma}{d}\right)}{\sin\left(\frac{\pi\gamma}{d}\right)}-1\right].
\label{eq:AD_joint_prob}
\end{eqnarray}
To apply this joint probability into the CGLMP inequality in Eq.~(\ref{Bell}), we follow  the definition  introduced in Ref.~\cite{CGLMP}:
\begin{eqnarray}
P(A_s = B_t + k) = \sum_{n=0}^{d-1} P(A_s = n, B_t = n+ k \mod d).\nonumber \\
\end{eqnarray}
Also  due to the symmetrical structure,  the  probability can be computed using simplified definition: 
\begin{eqnarray}
P_{QM}^{AD} (A_s=B_t+k) &=& \sum_{n=0}^{d-1} P_{QM}^{AD}(A_s = n, B_t = n+ k) \nonumber \\
&=& d P_{QM}^{AD} (A_s = k, B_t = 0).
\end{eqnarray}
This leads to  the following relation:
\begin{eqnarray}
P_{QM}^{AD} (A_1=B_1+k) &=& P_{QM}^{AD} (B_1=A_2+k+1) \nonumber \\
&=& P_{QM}^{AD} (A_2=B_2+k) \nonumber \\
&=& P_{QM}^{AD} (B_2=A_1+k).
\end{eqnarray}
Finally, for amplitude damping noise the quantum value of CGLMP inequality in Eq.~(\ref{Bell}) reads
\begin{equation}
\label{quantum}
I^{AD}_{QM}(d, r)=4\sum_{k=0}^{\lfloor d/2 \rfloor-1} \left( 1-\frac{2k}{d-1} \right) \left[ q_{k} (r)-q_{-(k+1)} (r) \right],
\end{equation}
where $q_k (r)=P_{QM}^{AD} (A_1=B_1+k)$ and $\lfloor p \rfloor$ implies the integer part of $p$.

Now, we analyze the value of Eq.~(\ref{quantum}) for $d \to \infty$. The probability $q_k (r)$ reads
\begin{eqnarray}
&&q_k (r)=\frac{1-(d-1)(r-2)r}{d^2} +\frac{(1-r)}{d^2} \times \nonumber \\
&&\left[(1-r)\frac{\sin^2 \left(\frac{\pi(d-1)(k+1/4)}{d}\right)}{\sin^2\left(\frac{\pi (k+1/4)}{d}\right)}+\frac{\sin\left(\frac{\pi(2d-1)(k+1/4)}{d}\right)}{\sin\left(\frac{\pi(k+1/4)}{d}\right)}-1\right]. \nonumber
\label{eq:AD_qk_prob}
\end{eqnarray}
In the limit of $d \to \infty$, the only first sine function of $q_k (r)$ survives
%\begin{eqnarray*}
%\frac{1}{d^2}\frac{\sin^2 \left(\frac{\pi(d-1)(k+1/4)}{d}\right)}{\sin^2\left(\frac{\pi (k+1/4)}{d}\right)} \to \frac{1}{2 \pi^2 (k+1/4)^2},
%\end{eqnarray*}
and other terms reduce to zero. Finally, we have
\begin{eqnarray}
\label{infinity}
I^{AD}_{QM}(\infty, r)&=&\frac{2 (1-r)^2}{\pi^2}\sum_{k=0}^{\infty}\left[\frac{1}{(k+\frac{1}{4})^2}-\frac{1}{(k+\frac{3}{4})^2}\right] \nonumber \\
&=&(1-r)^2\frac{32 \times \text{Catalan}}{\pi^2} \nonumber\\ &\simeq & 2.9698(1-r)^2,
\end{eqnarray}
where Catalan's constant is Catalan $\simeq 0.9159$.

\end{document}